\documentclass[aps, preprint]{revtex4}
\usepackage{graphicx}
\usepackage{epsf}
\usepackage{wrapfig}
\usepackage{epsfig}
\def\bk{{\mbox{\boldmath$k$}}}
\def\bq{{\mbox{\boldmath$q$}}}

\def\bp{{\mbox{\boldmath$p$}}}

 \def\br{{\mbox{\boldmath$r$}}}
   
   \def\bP{{\mbox{\boldmath$P$}}}
   \def\bp{{\mbox{\boldmath$p$}}}

\def\br{{\mbox{\boldmath$r$}}}
\def\b0{{\mbox{\boldmath$0$}}}

\def\bk{{\mbox{\boldmath$k$}}}
\def\bq{{\mbox{\boldmath$q$}}}

\def\bp{{\mbox{\boldmath$p$}}}
\def\boldDelta{{\mbox{\boldmath$\Delta$}}}

\def\br{{\mbox{\boldmath$r$}}}
\def\b0{{\mbox{\boldmath$0$}}}

\newcommand{\ra}{\,\rangle}
\newcommand{\la}{\,\langle}

\def \b #1{ {\bf #1}}
\newcommand{\be}{\begin{eqnarray}}
\newcommand{\ee}{\end{eqnarray}}

\def \b #1{ {\bf #1}}

\def \b #1{ {\bf #1}}

     \font\tenbifull=cmmib10 scaled 1200 
     \font\tenbimed=cmmib9
     \font\tenbismall=cmmib7
       \def\bmit{\fam9 }
\mathchardef\bbkappa="7114
\mathchardef\bbrho="711A
\mathchardef\bbsigma="711B
\mathchardef\bbtau="711C
\mathchardef\bbvarrho="7125
\mathchardef\bbvarsigma="7126
\mathchardef\bbxi="7118
\def\boldkappa{{\bmit\bbkappa}}
\def\boldrho{{\bmit\bbrho}}

\begin{document}
\vskip 2mm \date{\today}\vskip 2mm
\title{A non factorized calculation of the process
  $\bf ^3He(e,e'p)^2H$ at medium energies}
\author{C. Ciofi degli Atti}
\author{L.P. Kaptari}
\altaffiliation{On leave from  Bogoliubov Lab.
      Theor. Phys.,141980, JINR,  Dubna, Russia}
\address{Department of Physics, University of Perugia and
      Istituto Nazionale di Fisica Nucleare, Sezione di Perugia,
      Via A. Pascoli, I-06123, Italy}
 \vskip 2mm

\begin{abstract}
\vskip 5mm
The exclusive  process
$^3He(e,e^\prime p)^2H$
has been analyzed using realistic few-body wave functions corresponding to the $AV18$ interaction
 and treating
 the final state interaction (FSI)  within  the  Eikonal Approximation to describe
 the multiple rescattering
 of the struck nucleon with the nucleons of  the spectator two-nucleon system.
Calculations have been performed in  momentum space so that
the nucleon electromagnetic current could be left  in  the fully covariant form
 avoiding by this way non relativistic reductions and  the factorization  approximation.
The results of calculations, which   are compared with recent JLab  experimental data,
 show that the left-right asymmetry  exhibit a clear dependence upon
 the multiple scattering in the final state and demonstrate the breaking down of the
factorization approximation at $\phi=0$ i.e. for  "negative" and large $\geq 300MeV/c$
 values of the missing momentum.
\end{abstract}
\maketitle

 \section{Introduction}
Recent experimental data from Jlab  on exclusive electro-disintegration of
$^3He$~\cite{jlab1,benmokhtar}
are at present the object of intense theoretical activity (see
\cite{nashPR,nashPRL,Schiavilla,laget4,laget2,laget5} and References therein).
 In Refs. \cite{nashPR,nashPRL}   the 2- and 3-body break up  channels have been calculated
 within an approach where:

i) initial state
  correlations (ISC) have been taken care of  by the use of  the status-of-the-art
  few-body wave functions~\cite{pisa} corresponding to the $AV18$
  interaction \cite{av18};

ii) final state interactions (FSI) have been
treated by a  Generalized Eikonal Approximation~\cite{mark}, which
represents an   extended Glauber  approach (GA) based upon
 the evaluation of the relevant
  Feynman diagrams that describe  the rescattering of the
   struck nucleon in the final state, in analogy with the
  Feynman  diagrammatic approach  developed for the treatment
  of elastic hadron-nucleus scattering  \cite{gribov,bertocchi}.

 In \cite{nashPR,nashPRL}  theoretical calculations
 have been compared with  preliminary Jlab data
 covering a region of "right " ($\phi=\pi$, $\phi$ being the azimuthal  angle of the detected
 proton, with respect to the
 momentum transfers $\bq$) values of the missing momentum
 $ p_m \leq\,1.1 \, \, GeV/c$ and missing energy $E_m\,
 \leq 100 \,MeV$. Published data  \cite{jlab1}, however,
 cover  both the right  ($\phi=\pi$ and $p_m \leq 1.1\,\, GeV/c$) and
 left  ($\phi=0$ and  $p_m \leq  0.7\,\, GeV/c$ )
 values of the missing momentum  which  have not been  considered in
  \cite{nashPR,nashPRL}. It is the  aim of this paper  to analyze the process in the
  entire kinematical range improving,  at the same time,  our theoretical approach.
As a matter of fact,  previous calculations of ours, which took into account the Final
 State Interaction (FSI),
 have been  based upon  the factorization
approximation  which, as is well known, leads to a form of the cross section in terms of
 a product of two factors,
one describing the
electromagnetic electron-nucleon interaction, the other depending
 upon   nuclear structure and the strong interaction of nucleons in the final state.
 The factorization form is exactly satisfied  in the Plane Wave Impulse Approximation (PWIA),
  but it is however  violated
 in presence of FSI effects.

  Within the factorization approximation, the $\phi$-dependence
  of the  cross section
   is only due to  the $\phi$-dependence
  of the  elementary
  cross section for  electron scattering off  a moving nucleon \cite{forest}.
  Such a dependence is a very
  mild one and the recent data\cite{jlab1} on the left-right asymmetry unambiguously
  demonstrates that at  $p_m\geq\, 0.35 \, \, GeV/c$,
  the  cross section at $\phi=0$  appreciably differs from the one at  $\phi=\pi$.
  This, as is well known,   is  clear evidence
   that
  the factorization approximation cannot explain the left-right asymmetry.
  Several non factorized
  calculations
  appeared
  in the past. It should however be pointed out that most of them worked in configuration
  space, and in so doing the on mass shell current operator, which is exactly defined in
  momentum space, had
  to be reduced non relativistically  by different prescriptions. In the present paper
  we extend our approach by releasing  the factorization approximation and, at the same time,
   avoiding  non relativistic reductions
   by directly performing our calculations in momentum space,  treating the
  full current operator without any approximation.
  The $^3He$ wave function  of the
   Pisa group \cite{pisa}, corresponding to the AV18
   interaction \cite{av18} is  used   in the calculations.
   We   do not consider, for the time being,
   Meson Exchange Currents (MEC), $\Delta$-Isobar Configurations, and similar effects,
    which have been the object of intensive  theoretical studies in $A(e,e'p)B$
    processes off both few-body systems (see e.g. \cite{laget5,vanleuwe})
     and complex nuclei (see e.g. \cite{ryckebusch} and References therein quoted).
     We
   fully concentrate on the effects of the FSI, treating the initial and final state correlations,
   the Final State Interaction and the current operator within a parameter-free self-consistent
    approach.

   Recently \cite{Schiavilla}, the $^3He(e,e'p)^2H$ process and the left-right asymmetry
   have  been calculated within a non factorized
    GA approach, considering also the effects of
   MEC, adopting a non relativistic form for the nucleon electromagnetic current operator.

\section{The process $^3He(e,e'p)^2H$. Basic formulism}
 \label{sec:2}

We will consider  the process
\be
e+A=e'+p + (A-1)_f
\label{reac}
\ee
 where the relevant kinematical variables are defined as follows:
$k=(E,\b{k})$  and
 $k^{'}=(E^{'},{\b{k}}^{'})$, are  electron momenta before and after interaction,
 $P_A=(E_A,{\b{P}}_A)$ is  the momentum of the target nucleus,  $p_1=(\sqrt{{{\b{p}}_1}^2 +m_N^2},
 {{\b{p}}_1})$ and $P_{A-1}=(\sqrt{\b{P}_{A-1}^2
  +(M_{A-1}^{f})^2},{\b{P}}_{A-1})$,
   are the momenta
  of the final
 proton and the final
 $A-1$ system,
   $m_N$ is  the nucleon mass,
   $M_{A-1}^{f}=M_{A-1}+E_{A-1}^f$, where $E_{A-1}^f$ is the {\it intrinsic}
    excitation energy of the $A-1$ system. The
4-momentum transfer is $Q^2\equiv -q^2=(\nu,\bq)$.
The  relevant  quantities which characterize
the process are
   the
{\it missing momentum}\,\, ${\b p}_m$  (i.e.   the total momentum of the $A-1$ system),
and  the\,\, {\it missing energy} \,\, $E_m$ defined, respectively, by
\be
{\b p}_m = {\b q} - {{\b p}_1} \,\,\,\,\,\,  E_m=
\sqrt{P_{A-1}^2}+m_N -M_A \,\, = E_{min} + E_{A-1}^f .
\label{missing}
\ee
where $E_{min} = M_{A-1} +m_N - M_A =|E_A| - |E_{A-1}|$ is the threshold energy for the two-body break-up (2bbu)
channel.
  The differential cross section
for the exclusive process  has the  following form
\be
{d^6 \sigma \over  d \Omega ' d {E'} ~ d^3{\b p}_m} =
\sigma_{Mott} ~ \sum_i ~ V_i ~ W_{i}^A( \nu , Q^2, {\b p}_m, E_m),
\label{2}
 \ee
where
$i \equiv\{L, T, TL, TT\}$, and $V_L$, $V_T$, $V_{TL}$, and $V_{TT}$ are well-known
kinematical factors~\cite{electron};
the nuclear response functions $W_i^A$ are
\be &&
W_L=\left[ \frac{\bq^2}{Q^2}\,  W_{00} \right ];
\qquad W_{TL}\cos \phi = \frac{|\bq|}{\sqrt{Q^2}}\left[ 2\Re \left(W_{01}-W_{0-1}
\right)\right];\nonumber
\\[2mm]&&
W_T=\left[ W_{11} +W_{-1-1}\right];\qquad
W_{TT}\cos 2\phi =\left( 2\, \Re (W_{1-1}) \right),
\label{responses}
\ee
with
\be &&
W_{\lambda\lambda'}=(-1)^{\lambda+\lambda'}\varepsilon_\lambda^\mu W_{\mu\nu}
\varepsilon_{\lambda'}^{*\ \nu}
\label{munu}
\ee
 $\varepsilon_\lambda$ being the   polarization vectors of the
 virtual photon. The hadronic tensor
$W_{\mu\nu}^{A}$ is defined as follows
\begin{widetext}
\begin{eqnarray}
W_{\mu\nu}^{A} & = & \frac{ 1}{4\pi M_A} {\overline
{\sum_{\alpha_A}}} \sum_{ \alpha_{A-1}, \alpha_N}
(2\pi)^4 \delta^{(4)} (P_A + q - P_{A-1} -p_1)\times
\nonumber\\
 &\times&\langle \alpha_A \b P| {\hat J_\mu^A(0)} |
\alpha_N{{\b p}_1} , \alpha_{A-1}{\b P}_{A-1}  E_{A-1}^f \rangle
\langle  E_{A-1}^f  {\b P}_{A-1} \alpha_{A-1},\b p_1  \alpha_N  | {\hat J_\nu^A(0)} |
\alpha_A\b P_A\rangle ~,
\label{hadrontz}
\end{eqnarray}
\end{widetext}
where  $\alpha_i$ denotes the set of discrete quantum numbers of the systems
$A$, $A-1$ and the nucleon $N$ with momentum $\bp_1$. In Eq. (\ref{hadrontz})
  the vector $|\alpha_N {{\b p}_1}, \alpha _{A-1} {\b P}_{A-1}  E_{A-1}^f \rangle$
  consists asymptotically of the  nucleon $N$ and the  nucleus $A-1$,
  with momentum ${\b P}_{A-1}$ and intrinsic excitation
  energy $E_{A-1}^f$.
The evaluation of the nuclear response functions $W_{i}^A$ requires
the knowledge of  both the nuclear vectors
$|\alpha_A\b P_A\rangle$ and  $|\alpha_N{{\b p}_1} , \alpha_{A-1}{\b P}_{A-1}  E_{A-1}^f
\rangle$,  and the nuclear current
operators ${\hat J_\mu^A}(0)$.
In the present  paper we describe  the two- and
three-body ground states in terms of realistic wave functions generated by
modern two-body interactions~\cite{pisa}, and
  treat the final state interaction by a diagrammatic approach of  the
elastic rescattering of the struck nucleon with the nucleons of the $A-1$
system \cite{nashPRL, misak,nashPR}.
We consider the interaction of the incoming virtual photon $\gamma^*$
with a bound nucleon (the active nucleon)
of low virtuality ($p^2\sim m_N^2$) in the
  quasi-elastic  kinematics i.e.  corresponding to $x\equiv Q^2/2m_N\nu\sim 1$. In the
 quasi-elastic kinematics, the virtuality  of the struck nucleon after
    $\gamma^*$-absorption  is also rather low and,
    provided ${\b p}_1$  is sufficiently high, nucleon rescattering
    with the "spectator" $A-1$  can be   described to a large extent in terms of  multiple
    elastic scattering processes in the
    eikonal approximation~\cite{nashPRL, misak,nashPR}. It should be pointed out that
    even within such an approximation one encounters problems in treating
    the operator of the electromagnetic current for off-mass shell nucleons.
Up to now most approaches to the  process (\ref{reac}) for
    complex nuclei, were based upon a non relativistic reduction of
    the on mass-shell nucleon current operator $\hat j_\mu$
    (the Foldy-Wouthuysen transformation)
    with subsequent, non relativistic, evaluations  of matrix elements in
    co-ordinate space. In principle, the non relativistic
    reduction can be avoided by using  the fully covariant expressions for the current
    operator $\hat j_\mu$ within
    the factorization approximation (FA)  or by performing calculations in momentum space.
    In latter case,
    calculations  for complex nuclei  in  momentum space are hindered
     by the fact that  realistic nuclear wave functions are obtained
    in co-ordinate space.  As for the factorization approximation,
    it should be considered it  not only guarantees that relativistic kinematics can be treated
    correctly, which is a prerequisite at high energies, but it also provides in various
    instances a
     satisfactory agreement with experimental data ~\cite{nashPRL}.
    However, the inadequacies of the FA clearly manifest themselves in the calculation
    of specific quantities such as, for example,
    the
    left-right asymmetry with respect to the azimuthal  angle $\phi$:  if factorization
    holds, this quantity
       must precisely follow the well known behavior of the corresponding asymmetry
    in the electron-nucleon elastic scattering~\cite{forest} so that deviations from such a
    behavior would represent   a stringent evidence  of the breaking down  of
    the  FA.

   In this paper the results of calculations  of the left-right asymmetry
  of the process $^3He(e,e'p)^2H$ obtained in the momentum space
  using realistic wave functions  will be presented.

\subsection {The Final state interaction}

In  co-ordinate space
the initial and final states  of the process under consideration have the following form
\be &&
\Phi_{^3He}(\br_1,\br_2,\br_3) ={\hat{\cal A}}  e^{i {\bf P R}}\Psi_3(\boldrho,\br),
\nonumber\\&&
\Phi_f^*(\br_1,\br_2,\br_3)={\hat{\cal A}} S(\br_1,\br_2,\br_3)e^{-i \bp^{\: \prime}\br_1}
e^{-i {\bf P}_D {\bf R}_D} \Psi_D^*(\br)
\label{states}
\ee
where ${\hat {\cal A}}$ denotes a proper antisymmetrization  operator and the $S$- matrix
describing the final state interaction of nucleons within the eikonal approximation is
\be
S(\br_1,\br_2,\br_3)=\prod_{j=2}^3  \left[ 1-\theta \left(\br_{j ||}-\br_{1||}\right)
\Gamma\left( \br_{j\perp}-\br_{1^\perp} \right )\right],
\label{smatr}
\ee
where the profile-function $\Gamma(\br_\perp)$ is defined as
\be
\Gamma(\br_\perp)=\frac{1}{2\pi i k^*}\int d^2 \boldkappa_\perp f_{NN}(\boldkappa_\perp)
e^{-i\boldkappa_\perp \br_\perp}
\label{profil}
\ee
and $\boldrho$, $\br$ and ${\bf R}$ are three-body  Jacobi  co-ordinates.
In Eq. (\ref{profil}) $f_{NN}(\boldkappa)$ is  the
  elastic scattering amplitude of two nucleons with  center-of-mass
momentum $k^*$.
By  approximating  the nuclear electromagnetic current operator
with a sum of nucleonic currents $\hat j_\mu(i)$ and supposing that the virtual photon
interacts with the nucleon "1",  one has
\be
 J_\mu^{A}=\int\prod d\br_i \Phi_f^*(\br_1,\br_2,\br_3)j_\mu(1)e^{-i {\bf qr_1}}
\Phi_{^3He}(\br_1,\br_2,\br_3).
\label{jmu}
\ee

 In what follows we consider the reaction (\ref{reac}) at
  relatively large (few $GeV/c$)
 momentum transfers, which implies large relative momenta of the
 particles in the final states. This allows one to safely neglect the
 spin-flip terms in the $NN$ amplitude considering only its central part.
Then the matrix element  (\ref{jmu}) can be  re-written  in the momentum space    as follows

\begin{widetext}
\be
 J_\mu^{A} =\sum_\lambda \int\frac{d^3p}{(2\pi)^3} \frac{d^3\kappa}{(2\pi)^3}
S(\bp,\boldkappa)
\la s_f|j_\mu(\boldkappa-\bp_m;\bq)|\lambda\ra
{\cal O} (\bp_m-\boldkappa,\bp; {\cal M}_3,{\cal M}_2,\lambda),
\label{ja}
\ee
\end{widetext}
where the overlap integral ${\cal O} (\bp_m-\boldkappa,\bp; {\cal M}_3,{\cal M}_2,\lambda)$ is
defined by
\begin{widetext}
\be
\label{overl}
{\cal O} (\bp_m -\boldkappa, \bp; {\cal M}_3,{\cal M}_2,\lambda)=
\int d\boldrho d\br e^{i({\bf p}_m-\boldkappa)\boldrho}
e^{i\bp\br/2}\Psi_3(\boldrho,\br)\Psi_D^*(\br)\chi_{\frac12 \lambda}^+
\ee
\end{widetext}
and the Fourier-transform of the eikonal $S$-matrix is
\be
S(\bp,\boldkappa)
=\int d\br d\boldrho e^{-i\bp \br} e^{i\boldkappa \boldrho}
S(\boldrho,\br).
\label{simp}
\ee

The quantities ${\cal M}_3$, ${\cal M}_2$ and  $s_f$  represent the
projections of the angular momentum of  $^3He$, the deuteron and the final proton,
 respectively, and
$\lambda$  denotes the spin projection of the
proton  before the absorbtion of the virtual photon.

By considering different terms in the $S$-matrix (\ref{smatr}),
we are in the  position to calculate different contributions
(PWIA and  single and double rescattering) in  the nuclear matrix elements
$J_\mu^A$, eq. (\ref{jmu}), and in  the response functions $W_i$, eq. (\ref{responses}).

 1. {\it The  PWIA}

In absence of  FSI the $S$-matrix (\ref{smatr}) in  co-ordinate space is
$S(\br_1,\br_2,\br_3)=1$ and, correspondingly,
$S(\bp,\boldkappa)=(2\pi)^6\delta^{(3)}(\bp)\delta^{(3)}(\boldkappa)$. This allows one
to recover
 the well-known expression for the electromagnetic current
(\ref{jmu}) in terms of the Fourier transform of an overlap integral of the wave functions in
 co-ordinate space
\begin{widetext}
\be
J_\mu^A(PWIA)=\sum_\lambda \la s_f|j_\mu(-\bp_m;\bq)|\lambda\ra
\int d\boldrho e^{i\bp_m\boldrho}\int d\br
\Psi_{{\cal M}_3}(\boldrho,\br)\Psi_{{\cal M}_2}^*(\br)\chi_{\frac12 \lambda}^+.
\label{pwia}
\ee
\end{widetext}

\begin{figure}[!htp]
\epsfig{file=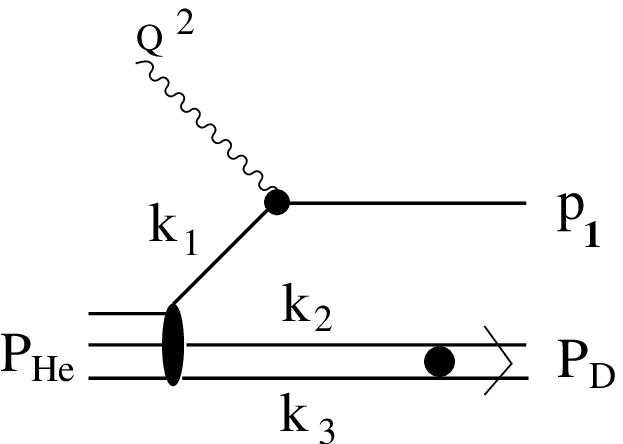,width=6cm,height=4cm}
\caption{ The Feynman diagram  for the
the process $^3He(e,e^\prime p)^2H$ in plane wave impulse approximation (PWIA).}
      \label{fig1}
\end{figure}

Equation (\ref{pwia}) corresponds  exactly  to the Feynman diagram shown in Fig.~\ref{fig1}.
Note  that the square of the matrix element (\ref{pwia}), averaged over initial
 (${\cal M}_3$) und summed over final (${\cal M}_2$ and $s_f$) spin projections,
 is  diagonal  with respect to the summation indices
$\lambda,\lambda'$ (see, e.g., Ref. \cite{nashPR}), so that in the
 response functions and, consequently,  the
 cross section, factorize in the well known form  in terms of the
  familiar spectral function \cite{ciofiSpectral} and
   the electron-nucleon cross section, $\sigma_{eN}$~\cite{forest}.

2. {\it Single rescattering.}

\begin{figure}[!htp]
\epsfig{file=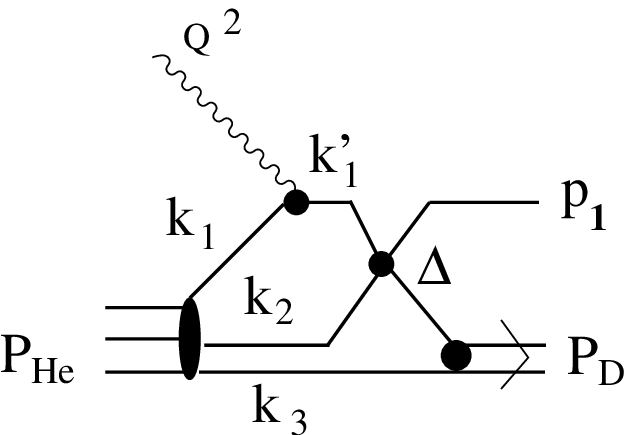,width=6cm,height=4. cm}
\caption{ Single rescattering  diagram  for the
the process $^3He(e,e^\prime p)^2H$. The missing momentum
$\bp_m$ is defined as $\bp_m=\bP_D$.
 The momentum of the active proton $\bk_1$ before
 the electromagnetic interaction satisfies the
 relation $\bk_1= -(\bk_2+\bk_3)=-\bp_m+\boldDelta $}
      \label{fig2}
\end{figure}

The corresponding part of the $S$-matrix for the single rescattering process  is

\be
S(\bp,\boldDelta)=-\frac{(2\pi)^4}{k^*}
\frac{f_{NN}(\boldDelta_\perp)}{\boldDelta_{||}-i\varepsilon}
\left[ \delta \left( \bp -\frac{\boldDelta}{2}\right) +
\delta \left( \bp +\frac{ \boldDelta}{2}\right)\right],
\ee
which leads to
\begin{widetext}
\be && \!\!\!\!\!\!\!\!\!\!\!\!\!\!\!\!
J_\mu^{A(1)}=\sum_\lambda  \int \frac{d\boldDelta}{(2\pi)^2 k^* }
\la s_f|j_\mu(\bk_1;\bq)|\lambda\ra \frac{f_{NN}(\boldDelta_\perp)}{\boldDelta_{||}-i\varepsilon}
\times\nonumber\\&&
\left[
{\cal O} (-\bk_1,\boldDelta/2; {\cal M}_3,{\cal M}_2,\lambda)
+{\cal O} (-\bk_1 ,-\boldDelta/2; {\cal M}_3,{\cal M}_2,\lambda)
\right],
\label{single}
\ee
\end{widetext}
where $\bk_1$ is the momentum of the
proton   before $\gamma^*$ absorption, $\bk_1=\boldDelta-\bp_m$, and
 $\boldDelta$ is the momentum transfer
in the $NN$ interaction. The corresponding  Feynman diagram is depicted in Fig.~\ref{fig2}.
The longitudinal part of the nucleon propagator can be computed  using
the relation
\be
\frac{1}{\boldDelta_{||}\pm i\varepsilon}=\mp i\pi\delta(\boldDelta_{||}) + P.V.\frac{1}{\boldDelta_{||}}.
\label{pv}
\ee
 It should be pointed out that in the eikonal approximation  the trajectory of
 the fast  nucleon is a straight line so that all
 the "longitudinal" and "perpendicular" components are defined
 in  correspondence to this trajectory, i.e., the $z$ axis in our case
 has  to be directed along the   momentum of the detected fast proton.
It can also be seen  that since the argument of the nucleonic current
$\la s_f|j_\mu(\bk_1;\bq)|\lambda\ra$
 is related to the
integration variable $\boldDelta$, the factorization form is no longer fulfilled.
However, as shown in Ref. \cite{nashPR}, if in the integral (\ref{ja}) the longitudinal part
can be neglected, the factorization form can be approximately recovered.

In actual calculations the elastic amplitude $f_{NN}$ is usually parametrized in the
following  form
\be
f_{NN}(\boldDelta_\perp)=k^* \frac{\sigma^{tot}(i+\alpha)}{4\pi}e^{-b^2\boldDelta_\perp^2/2},
\label{slope}
\ee
where the slope parameter $b$, the total nucleon-nucleon cross section $\sigma^{tot}$ and
the ratio  $\alpha$ of the real to the imaginary parts  of the forward scattering amplitude,
are taken from experimental data.

3. {\it Double rescattering.}

\begin{figure}[!htp]
\hspace*{-2mm}                     
\epsfig{file=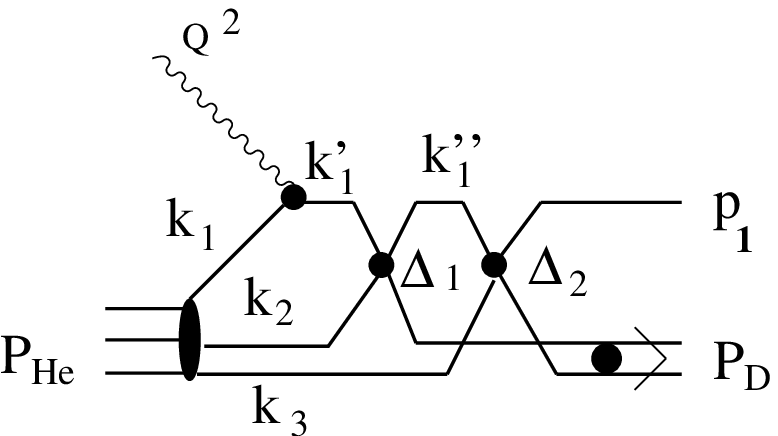,width=6cm,height=4. cm}
\caption{ Double rescattering  diagram  for the
the process $^3He(e,e^\prime p)^2H$. The missing momentum
$\bp_m$ is defined as $\bp_m=\bP_D$.
 The momentum of the active proton $\bk_1$ before
 the electromagnetic interaction satisfies the
 relation $\bk_1= -(\bk_2+\bk_3)=-\bp_m+\boldDelta_1+\boldDelta_2 $}.
      \label{fig3}
\end{figure}

In the same manner  the double rescattering $S$-matrix can be obtained in the following form
\begin{widetext}
\be
\label{double}
S(\bp,\boldkappa)=-\frac{(2\pi)^2}{k_1^*k_2^*}\int
d\boldDelta_1 d\boldDelta_2
\frac{f_{NN}(\boldDelta_{1\perp})f_{NN}(\boldDelta_{1\perp})}
{\left( \boldDelta_{1||}+i\varepsilon\right)
\left( \boldDelta_{2||}+i\varepsilon\right)}
\delta\left( \bp + \frac{ \boldDelta_1-\boldDelta_2}{2}\right)
\delta\left( \boldkappa +  \boldDelta_1+\boldDelta_2 \right),
\label{sdoub}
\ee
\end{widetext}
and, correspondingly,  for the electromagnetic current one has
\begin{widetext}
\be &&
J_\mu^{A(2)}=\frac{1}{ (2\pi)^4k_1^*k_2^*}\sum_\lambda \int d\boldDelta_1 d\boldDelta_2
\frac{f_{NN}(\boldDelta_{1\perp})f_{NN}(\boldDelta_{1\perp})}
{\left( \boldDelta_{1||}+i\varepsilon\right)
\left( \boldDelta_{2||}+i\varepsilon\right)} \times\nonumber\\&&
\la s_f|j_\mu(\bk_1;\bq)|\lambda\ra
{\cal O} (-\bk_1,(\boldDelta_1-\boldDelta_2)/2; {\cal M}_3,{\cal M}_2,\lambda),
\label{dbl}
\ee
\end{widetext}
where now the proton momentum before interaction is $\bk_1= \boldDelta_1+\boldDelta_2-\bp_m$.
As in the previous case, $\boldDelta_{1,2}$ are the momentum transfers in $NN$ rescattering,
as depicted in Fig.~\ref{fig3}.

It can be  seen from   Eqs.~(\ref{ja}) and (\ref{dbl}) that
the matrix element of the nucleon current operator $\la s_f|j_\mu(\bk_1;\bq)|\lambda\ra$
is evaluated in  momentum space.
In the case of  on-mass-shell nucleons the corresponding expression is
\begin{widetext}
\be
\la s_f |j_\mu(\bk_1;\bq)|\lambda\ra=
\bar u(\bk_1+\bq,s_f)\left[ \gamma_\mu  F_1(Q^2)
+i\frac{\sigma_{\mu\nu}q^\nu}{2m_N} F_2(Q^2)\right]
u(\bk_1,\lambda),
\label{cc2}
\ee

or, due to  the Gordon identity,

\be
\la s_f| J_\mu (\bk_1,\bq)|\lambda\ra=
\bar u(\bk_1+\bq,s_f)\left[ \gamma_\mu \left( F_1(Q^2) + F_2(Q^2)\right )-(2k_1+q)_\mu F_2(Q^2) \right]
u(\bk_1,\lambda),\nonumber\\
\label{cc1}
\ee
\end{widetext}
where  $F_{1,2}(Q^2)$ are the Dirac and Pauli  nucleon form factors.
Eqs. (\ref{cc2}) and (\ref{cc1}) for on mass shell
nucleons are completely equivalent, however for the off mass shell case
 they could be rather
 different,  for, in this case the Gordon identity does not hold. This leads to some arbitrariness
and discussions about the actual choice of the nucleon current.
In our calculations,
following the de Forest prescription \cite{forest},
we adopt the nucleonic current in form of Eq.~(\ref{cc1}), usually referred to as
the "CC1" prescription.

\section{Results of calculations}

We have used the described formalism  to calculate the cross sections of the processes
(\ref{reac}).  All two- and three-body wave functions were taken to be
solutions of the  non relativistic Schr\"odinger equation with
the AV18 potential Ref.~\cite{pisa}. Calculations have been performed in PWIA and including
the full rescattering within the  eikonal approximation corresponding to the diagrams shown in
  Figs.~\ref{fig1}-\ref{fig3}.

\begin{figure}[!htp]
\centerline{
\hspace*{-2mm}                     
\includegraphics[scale=0.8 ,angle=0]{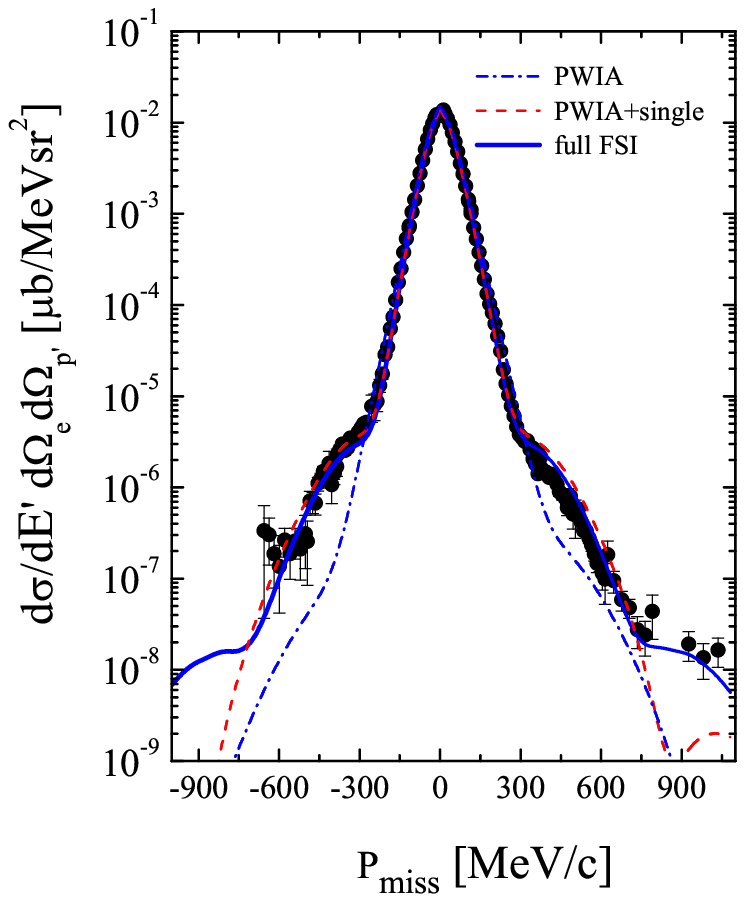}
\includegraphics[scale=0.83,angle=0]{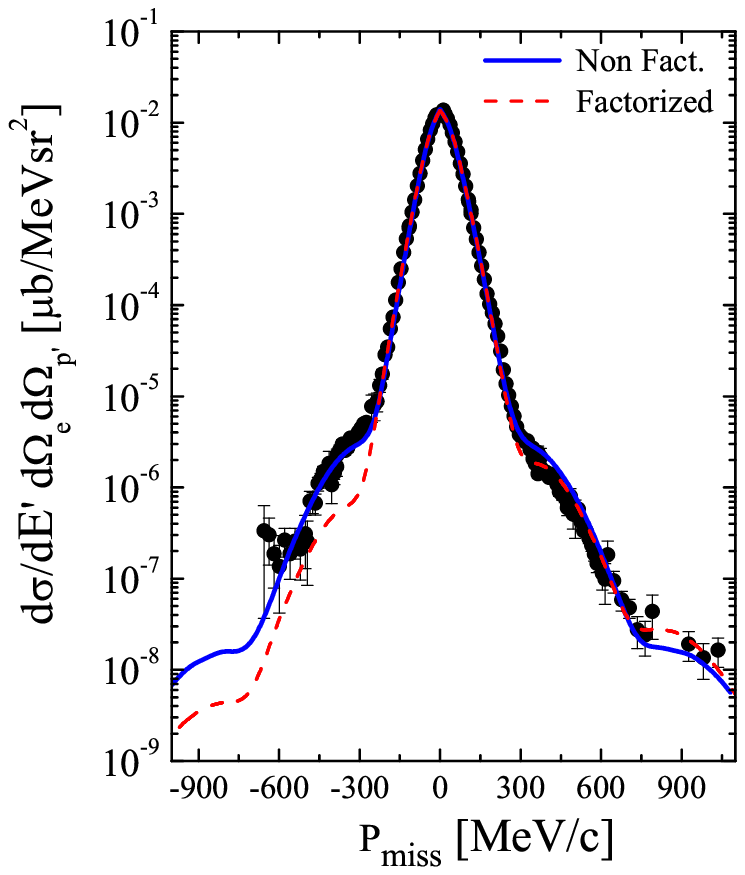}
}
 \caption{
The differential cross section for the process $^3He(e,e^\prime p)^2H$. In the left panel
the results of the non factorized calculations are shown.  {\it Dot-dashed curve}:
 PWIA; {\it dashed curve}
PWIA plus single rescattering FSI; {\it full curve}: PWIA plus single and double rescattering FSI.
In the right panel  the present non factorized
results ({\it full curve}) are compared with the results obtained within the
factorization ({\it dashed curve}). Experimental data from
ref. \cite{jlab1}}
      \label{fig4}
\end{figure}

The results of our calculations are shown in
  in  Fig. \ref{fig4} where they are compared
with  recent experimental data ~\cite{jlab1} corresponding to  $\phi=0$
( negative values of the missing momentum) and
$\phi=\pi$ (positive values of the missing momentum).
The relevant kinematical variables in the experiment were
 $|\b{q}|=1.5\,\, GeV/c$, $\nu=0.84\,\, GeV$, $Q^2=1.55 \, (GeV/c)^2$,
 and  $x\approx 1$.
In PWIA  the cross section is directly proportional to the two-body spectral function of  $^3He$.
 It  can be seen
 that  up to
 $|{\b p}_m|\sim 400\,\, MeV/c$,   PWIA and FSI results are almost the same and fairly
 well agree with the experimental data, which means  that
 the 2bbu process    $^3He(e,e'p)^2H$ does provide information on the two-body spectral function;
 on the contrary,
 at larger values of $|{\b p}_m|\geq 400\,\, MeV/c$  the PWIA
 appreciably underestimates the experimental data. It is however very
 gratifying to see that   when FSI is taken into account,  the disagreement
 is completely removed and an overall   good agreement
 between theoretical predictions and experimental data is obtained.
 It should be pointed out  that  at large missing momenta
 the experimental data shown in Fig. \ref{fig4}
 correspond to the  perpendicular kinematics,
 when  the deuteron momentum   is always
 almost perpendicular to the momentum of the final proton; in such
a kinematics the effects from  FSI are maximized, whereas in the so-called
parallel kinematics, they are minimized
 (see, e.g. ~\cite{mark}, ~\cite{niko}, ~\cite{mor01}).

 Fig.~\ref{fig4}
shows, however, that  in some regions  quantitative disagreements with data still exist.
Particularly  worth being mentioned  is  the disagreement in the region around
 $|{\b p}_m|\simeq 0.6- 0.65\,\, GeV/c$ at $\phi=0$.
 Other possible mechanisms
  in this kinematical range (MEC, $\Delta$ \cite{Schiavilla,laget4,laget2,laget5})
  which could remove this disagreement will be the object of  future investigations.

  We would  like to stress, that in our calculations no approximations have
  been made in the evaluation
  of the single and double scattering contributions to the FSI: proper
  intrinsic coordinates have been used and the energy dependence of the profile
  function  has been taken into account in the properly chosen CM system of the
  interacting pair. Note also, that the numerical values of the parameters
  are exactly the same for the
    left and right shoulders in the Fig. \ref{fig4}. The obtained
    results are clear evidence that the   difference
    in the "left" and "right" cross sections has  a dynamical origin  entirely
    governed by   FSI  effects.

\begin{figure}[!htp]
\hspace*{-2mm}                     
\epsfig{file=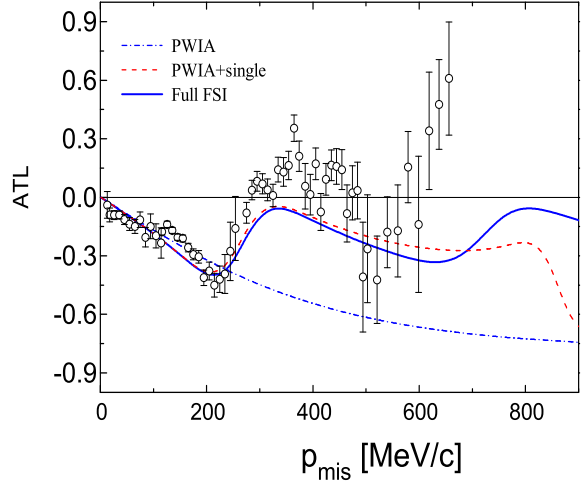, width=9cm,height=8 cm}
\caption{ The left-right asymmetry for the process
$^3He(e,e^\prime p)^2H$. {\it Dot-dashed curve}: PWIA; {\it dashed curve}: PWIA plus
single rescattering FSI; {\it full curve}: PWIA plus single and double rescattering FSI.
 Experimental data are from ref. \cite{jlab1}}
      \label{fig5}
\end{figure}

    The "left-right" asymmetry is defined as follows
\be
A_{TL} =\frac{d\sigma(\phi=0^o)-d\sigma(\phi=180^o)}{d\sigma(\phi=0^o)+d\sigma(\phi=180^o)}.
\label{atl}
\ee
 It can be seen from Eqs. (\ref{responses})
 that the numerator in (\ref{atl})  is proportional to $W_{TL}$, whereas the
 denominator does not contain $W_{TL}$ at all, i.e. the
 $A_{TL}$ is a measure of   the weight of the transversal-longitudinal
 components in the cross section,  relative to the other responses. For the elementary
 $eN$ cross section the behavior of the asymmetry $A_{TL}$ is known
  to be a negative and  decreasing function of the missing momentum ~\cite{forest}.
 It is clear that in the PWIA and within the FA  the
 asymmetry (\ref{atl}) for the process (\ref{reac}) must be exactly the same
 as in the $eN$ case. In Fig.~\ref{fig5}  the  asymmetry
 $A_{TL}$  for the process (\ref{reac}) computed within the present approach is shown
  together with the
 available experimental data~\cite{jlab1} . The
 dot-dashed line correspond to the
 PWIA,
 the dashed line includes  single rescattering FSI, and, eventually,
 the solid line includes the  full FSI. It can be  seen that  at $p_m \le 250 MeV/c$
  the PWIA result  is in  good agreement with the experimental
 data. However with increasing  $p_m$  the disagreement of the experimental data with the  PWIA
 predictions appreciably   increases.
 An interesting observation can be made from
 an inspection of the behavior of the asymmetry $A_{TL}$ in the region
 of $p_m$ corresponding to the  interference between different
 terms  of the rescattering $S-$matrix (cf. Fig.~\ref{fig4}).
 As a matter of fact, it can be  seen that in this region  the shape of the
 asymmetry, which  strongly depends upon the value of the missing momentum,
 exhibiting  a behaviour reflecting single and double rescattering
  in the final states. The change of slopes of the experimental data reflecting
  the multiple scattering structure has already been pointed out in Ref. \cite{nashPRL}.
  It is also interestring to note that, as in the case of other calculations \cite{Schiavilla},
 the theoretical
  asymmetry does not agree with the experimental data.
 It should be noted, however, that for values of $p_m$ up to $p_m\sim 650 MeV/c$
 the asymmetry is rather small $\sim 0-20\%$, i.e. the contribution
 of the  response function $W_{TL}$ to the total cross section
 is much smaller in comparison to other three responses,
 cf. Ref.~\cite{sabina1}. Correspondingly, at high values of the missing momentum
 the analysis of the asymmetry $A_{TL}$ does not allow one to
 draw definite  conclusions about the limits of validity of the FA.
For such a reason,  let us
define another quantity, which "amplifies" the limits of validity of the FA, namely
 following Ulmer et al. \cite{ulmer}, we  consider
the so-called reduced cross section $d\sigma_{red}$ defined by the ratio of the cross section
(Eq. (\ref{2})) to the electron nucleon  "CC1" cross section \cite{forest}, i.e.
\be
d\sigma_{red}=\frac{1}{\sigma_{cc1}}{d^6 \sigma \over  d \Omega ' d {E'} ~ d^3{\b p}_m}.
\label{reduced}
\ee
Then the deviation of the ratio
\be
R=\frac{d\sigma^{red.}(\phi=0)}{d\sigma^{red.}(\phi=\pi)}
\label{ratred}
\ee
from unity would be an indication of the breaking  down of the
FA. In Fig. \ref{fig5}  the ratio
(\ref{ratred}) calculated within the present approach is compared together
with the corresponding experimental quantity, obtained from data \cite{jlab1}. It can be
 seen that  up to values of $p_m\sim 0.3 GeV/c$ the FA
holds for both, $\phi=0$ and $\phi=\pi$ (cf. also the PWIA results in  Fig. \ref{fig4}).
At larger values of $p_m$ the ratio (\ref{ratred}) is larger than one,
with  a tendency to remain  constant as $p_m$ increases.

    \section{Summary and Conclusions}
\label{sec:4}
 We have calculated in momentum space the cross section of the
 processes  $^3He(e,e'p)^2H$, using realistic ground state two-and three-body
 wave functions and treating
 the FSI of the struck nucleon with the spectators
 within the  eikonal  approximation
 The method we have used is a very transparent
  and  parameter free  one: it is based upon
 Eqs. (\ref{ja}), (\ref{overl}), and (\ref{cc1},
  which only require the knowledge of
 the nuclear wave functions, since the FSI factor is fixed directly
 by  NN scattering data. At the same time, calculations are very involved
 mainly because of the complex structure of the wave function of Ref.~\cite{pisa},
which has to be firstly transformed to  momentum space and then used in calculations
of multidimensional integrals, including also the
computations of Principal Values (see eq. (\ref{pv})) together with
the Dirac algebra for the  electromagnetic current (\ref{cc1}).
Several aspects and results of our approach deserve the following comments:
\begin{enumerate}

\item our calculations have been performed in  momentum
       space with the electromagnetic current treated in a fully covariant
       form and with the factorization assumption released;
\item our approach does not rely on the factorization approximation;
\item the agreement between the results of our calculations and the
      experimental data for both $\phi=0$ and $\phi=\pi$,
      is a very satisfactory one, particularly
      in view the lack of freely adjustable parameter
      in our approach;
\item the violation of the factorization approximation is appreciable at negative values of
$\bp_m \geq 300 MeV/c$, whereas the non factorized and factorized results are in much better agreemen
in the whole range of positive values of $\bp_m$;

\item calculations of the 2bbu  disintegration channel  of  $^4He$, i.e.
   the process $^4He(e,e'p)^3H$, have already been
   performed \cite{hiko} within the factorization approximation
   using realistic wave functions and
   taking exactly into account  nucleon rescattering up to 3rd order.
   Calculations within a nonfactorized approach are in progress and will be reported
   elsewhere \cite{helium4};
   they should in principle
   yield results appreciably differing from
   the predictions based upon shell-model type four-body wave functions, thus allowing
   a study of NN correlations at densities comparable to the density of cold nuclei;

\item our results for $^3He$ generally agree with the ones obtained in Ref. \cite{Schiavilla},
     so that it would appear that the problem of the treatment of FSI at high
     values of $Q^2$ (or high ${\bf p}_1$) is under control;
\end{enumerate}

 \section{Acknowledgments}
The authors are  indebted to  A. Kievsky  for making available
the variational three-body  wave functions of the Pisa Group.
Thanks are due  to M.A. Braun  for stimulating discussions on the
 Feynman diagram approach to nucleon rescattering and to S. Gilad, H. Morita,  E. Piasetzky,
 M. Sargsian,
  R. Schiavilla and   M. Strikman for
 many useful discussions  concerning both the experimental and theoretical aspects of the topic
 considered in this paper.
L.P.K. is   indebted to  the University of Perugia and INFN,
Sezione di Perugia, for a grant and for warm hospitality.

.
\newpage
.

\vskip 2mm

\vskip 2mm

\vskip 2mm
\begin{figure}[!htp]
\hspace*{-2mm}                     
\epsfig{file=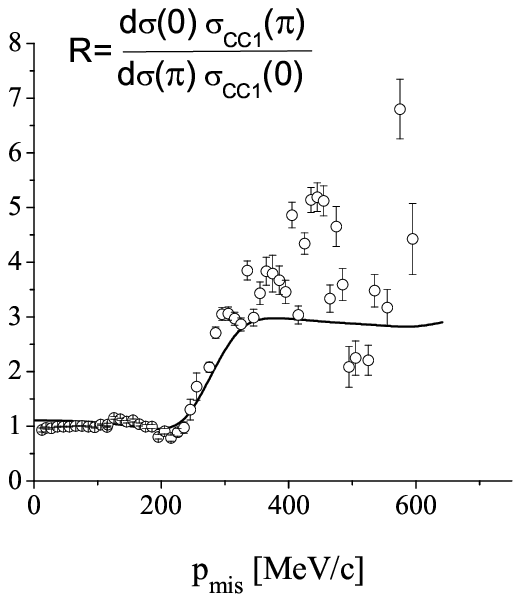,width=9cm,height=11 cm}
\caption{The reduced ratio (\ref{ratred}) obtained  within the present approach
({\it curve line}) compared  with  the corresponding experimental data \cite{jlab1}}
      \label{fig6}
\end{figure}


\begin{thebibliography}{99}
\bibitem{jlab1}
                  M. M.  Rvachev  et al.,  Phys. Rev. Lett. {\bf 94} (2005) 192302.
\bibitem{benmokhtar}
                 F. Benmokhtar et al., Phys. Rev. Lett. {\bf 94} (2005) 082305.
\bibitem{nashPR}
                C.Ciofi degli Atti, L.P. Kaptari, Phys. Rev. {\bf C71} (2005) 024005.
\bibitem{nashPRL}
                C.Ciofi degli Atti, L.P. Kaptari, Phys. Rev. Lett. {\bf 95} (2005) 052502.
\bibitem{Schiavilla}
           R. Schiavilla, O. Benhar, A. Kievsky, L.E. Marcucci,
           M. Viviani,  Phys. Rev. {\bf C72} (2005) 064003.
%
\bibitem{laget4}
                  J.-M. Laget, Few-Body Systems Suppl. {\bf 15} (2003) 171;
                   \texttt{nucl-th/0303052}.
\bibitem{laget2}
                   J.-M. Laget, Nucl. Phys. {\bf A579} (1994) 333.
\bibitem{laget5}
                  J.-M. Laget,  Phys. Rev. {\bf C72} (2005) 024001.
%
\bibitem{pisa}
                  A. Kievsky, S. Rosati and M. Viviani,
                  Nucl. Phys. \textbf{A551} (1993) 241 and  \textit{Private
                  communication}.
\bibitem{av18}
                  R. B. Wiringa, V. G. J. Stoks and  R. Schiavilla,
                  Phys. Rev. {\bf C51} (1995) 38.

\bibitem{mark}
                  L.L. Frankfurt, W.R. Greenberg, G.A. Miller, M.M. Sargsian, M.I.
                  Strikman, Z. Phys. \textbf{A352} (1995) 97.
%
\bibitem{gribov}
                  V.N.Gribov, Sov. Phys. JETP, {\bf 30} (1970) 709.
%
\bibitem{bertocchi}
                  L. Bertocchi, Nuovo Cimento, {\bf 11A} (1972) 45.
%
\bibitem{forest}
                  T. de Forest Jr., Nucl. Phys. {\bf A392} (1983) 232.
\bibitem{vanleuwe}
                  J. J. van Leeuwe {\it et al},  Phys. Lett.  \textbf{B523} (2001) 6.
%
\bibitem{ryckebusch}
                  S. Janssen, J. Ryckebusch, W. Van Nespen, D. Debruyne, Nucl.  Phys.
                  \textbf{A 672} (2000) 285.
\bibitem{electron}
                  S. Boffi, C. Giusti and F.D. Pacati, Phys. Rep. {\bf 226} (1993) 1.
\bibitem{misak}
                  M.M. Sargsian, Int. J. Mod. Phys.  \textbf{E10}(2001)405.
\bibitem{sabina}
                 S. Jeschonnek, Phys. Rev. {\bf C 63} (2001) 034609.
%
\bibitem{niko}
                  A. Bianconi, S. Jeschonnek, N. N. Nikolaev and  B. G. Zakharov,
                  Nucl. Phys. {\bf A608} (1996) 437.
\bibitem{mor01}
                  H. Morita, C.Ciofi degli Atti and  D. Treleani,
                  Phys. Rev.  \textbf{C60}(1999) 34603-1.
%
\bibitem{sabina1}
                 S. Jeschonnek, T.W. Donelly,
                 Phys. Rev. {\bf C57} (1998) 2438.
\bibitem{ulmer}
                  P.E. Ulmer, K.A. Aniol, H. Arenh\"ovel, J,-P. Chen, {\it et al.},
                  Phys. Rev. Lett., \textbf{89} (2002) 062301-1.
\bibitem{hiko}
            H. Morita, M. Braun, C. Ciofi degli Atti, D. Treleani,
             Nucl. Phys. {\bf A 699} (2002) 328
\bibitem{helium4}
                C.Ciofi degli Atti, L.P. Kaptari and H. Morita, {\it work in progress}.
\end{thebibliography}
\end{document}